\documentclass[acmlarge,nonacm]{acmart}
\AtBeginDocument{%
  }

\setcopyright{cc}
\setcctype{by}
\copyrightyear{2026}
\acmYear{2026}
\acmDOI{XXXXXXX.XXXXXXX}
\acmConference[AR On-the-Move — CHI '26]{the CHI 2026 Workshop on Next Steps for Augmented Reality On-the-Move: Challenges \& Opportunities}{April 16,
  2026}{Barcelona, Spain}

\usepackage{subcaption}

\begin{document}

\title{UI Placement as a Critical Design Factor for Augmented Reality During Locomotion}

\author{Pavel Manakhov}
\email{p.manakhov@lancaster.ac.uk}
\orcid{0000-0003-3443-4088}
\affiliation{
  \institution{Lancaster University}
  \city{Lancaster}
  \country{United Kingdom}
}

\author{Hans Gellersen}
\email{hwg@cs.au.dk}
\orcid{0000-0003-2233-2121}
\affiliation{
    \institution{Lancaster University}
    \city{Lancaster}
    \country{United Kingdom}
}
\affiliation{
    \institution{Aarhus University}
    \city{Aarhus}
    \country{Denmark}
}
\renewcommand{\shortauthors}{Manakhov \& Gellersen}

\received{12 February 2026}
\received[accepted]{15 March 2026}

\maketitle

\small
\textbf{ACM Reference Format:}

\noindent Pavel Manakhov and Hans Gellersen. 2026. UI Placement as a Critical Design Factor for Augmented Reality During Locomotion.
In \textit{Proceedings of the CHI 2026 Workshop on Next Steps for Augmented Reality On-the-Move: Challenges \& Opportunities (AR
On-the-Move — CHI ’26)}. ACM, New York, NY, USA, 4 pages. \url{https://doi.org/XXXXXXX.XXXXXXX}

\normalsize
\section{Introduction and Background}

Wearable Augmented Reality (AR), embodied in glasses and eventually contact lenses, represents the next frontier of computing interfaces. While the current generation of compact AR glasses, such as Xreal One and Viture Luma, tethered to portable consoles and laptops is transforming how gaming and stationary work is done in cafes and airplanes~\cite{Cheng25a}, future devices are anticipated to blur the boundaries between sedentary and mobile tasks. These systems will enable safe, efficient access to digital information and facilitate new forms of mobile productivity, such as walking meetings. This ability to switch between sedentary work and performing tasks on the go enhances flexibility and may promote a more active, healthier lifestyle~\cite{Chau10, Chang24}.

Being on the move, however, fundamentally redefines the spatial relationship between the AR interface and the user. In stationary contexts, a user interfaces (UI) affixed to the environment does not move much relative to a seated user. In contrast, the relative placement of a UI on the go can be affected by complex body movements, walking pace, and geometry of the environment, making it harder to perceive information from the UI and to select UI controls. \textbf{To design AR interfaces suitable for on-the-go use, we need to understand how UI placement — the spatial relationship between the user and the interface — affects interaction during locomotion}.

In our previous work, we focused on investigating how UI placement affects the perception of visual information and the accuracy of gaze pointing during physical locomotion. Fixations — periods during which the eyes remain relatively stable and aligned with an object of interest — are the fundamental building blocks of visual perception. For visual processing to be effective, the image of the object must be sufficiently stable on the retina~\cite{Borg15}. In “Gaze on the Go”~\cite{Manakhov24-GazeOnTheGo}, we studied the stability of fixations during linear locomotion as a function of object placement. This work compared visual acquisition of fully Head- and World-anchored virtual targets as baselines against HeadDelay and Path placements. In HeadDelay, targets followed head translation and rotation as in Head, but with a delay. In Path, targets floated in front of the walking participant at a fixed distance and height above the ground, aligning laterally with the user’s predicted path rather than with the user’s head.

\begin{figure}[t]
  \begin{minipage}{0.48\textwidth}
    \centering
    \includegraphics[width=0.75\linewidth]{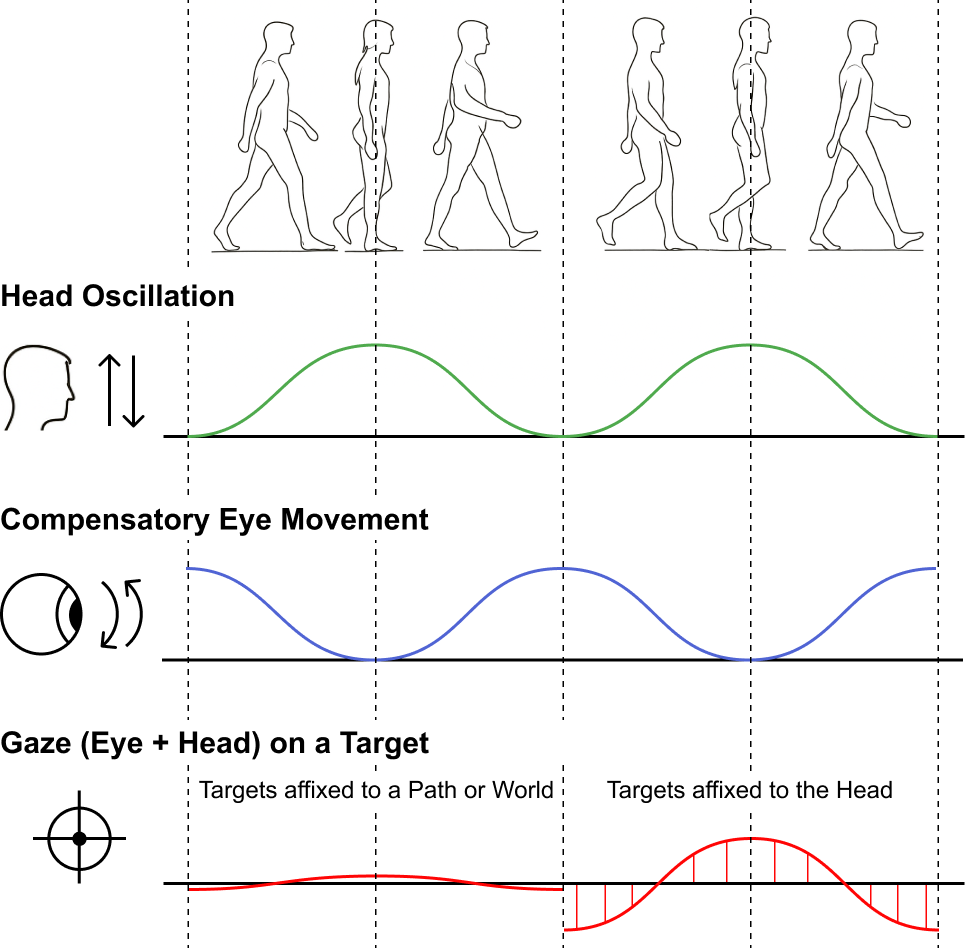}
    \caption{The head naturally oscillates laterally and vertically during motion. To counter these oscillations, humans rely on a number of vestibular- and visually-driven compensatory eye movements that are active during locomotion, with the vestibulo-ocular reflex (VOR) chief among them. These mechanisms can successfully counter head oscillations when the target is affixed to the path or world coordinates, as shown on the left. When the target is rigidly affixed to the head, it remains stationary within the field of view. Compensatory eye movements, however, continue to swing the gaze around the target, leading to reduced fixation stability, as shown on the right.}
    \label{fig:VOReffect}
  \end{minipage}
  \hfill 
  \begin{minipage}{0.48\textwidth}
    \centering
    \includegraphics[width=0.61\linewidth]{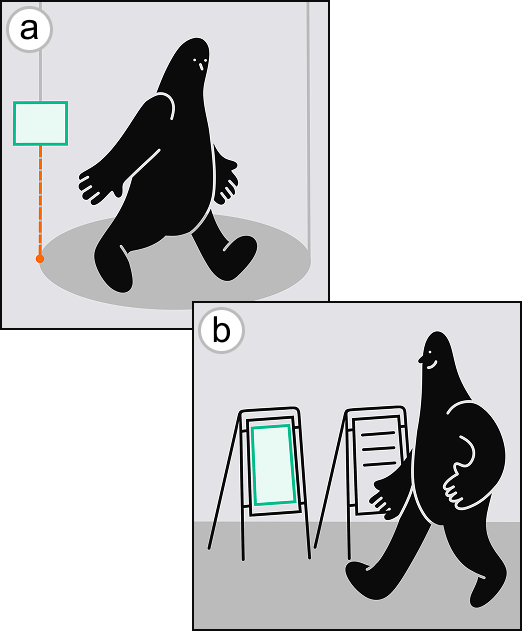}
    \caption{Less common types of mobile UI placement: (a) \textit{tag-along} UI placements define an area moving with the user, within which the UI remains stationary relative to the world. Once the interface hits the boundaries of this area, it is dragged along to remain within the user's field of displacement. This area can take the form of a vertical cylinder centered around the user~\cite{Lages_Bowman19-WalkingWithARWorkspaces, Belo22} or a view frustum~\cite{Microsoft21-Billboarding_TagAlong}; (b) UIs \textit{dynamically relocated in world coordinates} remain world-fixed until specific conditions are met. These conditions range from detection of physical billboards within the surroundings~\cite{Mann_Fung01} to overlaying UIs onto obstacles currently being fixated on by the user~\cite{Zhao24}.}
    \label{fig:nonTraditionalUIplacements}
  \end{minipage}
\end{figure}

The study results demonstrated that stabilizing targets in the plane perpendicular to the direction of locomotion aids visual perception, with Path and World behaving identically and achieving the highest fixation stability, Head performing the worst, and HeadDelay falling somewhere in between (\autoref{fig:VOReffect}). These findings directly inform how UIs should be presented in wearable AR. First-generation heads-up display (HUD) glasses, such as Meta Ray-Ban Display and Even Realities G2, which are designed for use in a wide range of contexts, including while on the move, would benefit from smoothing UI motion relative to the head using data from built-in inertial measurement units (accelerometers and gyroscopes). Future glasses with positional tracking would benefit from positioning virtual information using World and Path placements during locomotion.

Fixation stability also affects how accurately users can point at UI controls when gaze is used as an input modality on the go. Following the findings from \cite{Manakhov24-GazeOnTheGo}, in “Filtering on the Go” \cite{Manakhov24-FilteringOnTheGo}, we focused on improving gaze pointing accuracy by studying how online gaze filters perform under varying locomotion and UI placement conditions. Applying online filters smoothes the gaze signal, effectively minimizing its dispersion and, consequently, increasing pointing accuracy. Our computational experiment revealed that, given the nature of compensatory eye movements (\autoref{fig:VOReffect}), the gaze signal should be filtered differently depending on UI placement (e.g., suppressing low-frequency VOR movements increases pointing accuracy in Head but decreases it in Path- and World). Additionally, the study showed that filters are most efficient when applied to a gaze signal converted to the UI’s coordinate system. The practical implications of these findings render devices with no control over the gaze post-processing pipeline, such as Apple Vision Pro and Meta Quest Pro, less ideal for gaze-based interaction on the move. These results also inform the choice of online gaze filters and the parameters best suited for locomotion.

\section{Current Research Challenges \& Opportunities}

\paragraph{Interplay Between UI Placement and Interaction Techniques}

Human locomotion is a complex process involving translational and rotational oscillations of the upper body and head, which vary with walking speed. Regardless of input modality, be it direct touch or raycasting with the user’s hands, head, or eyes, these body movements significantly affect interaction performance. Importantly, however, the effect of movement on interaction is not direct; rather, it is mediated by the spatial placement of the UI relative to the user. For instance, when targets are affixed relative to the user’s walking direction at a fixed distance, raycast-based selection becomes slower as walking speed increases. In contrast, when targets are world-fixed, selection times decrease at higher walking speeds, as the targets appear larger when the user approaches them~\cite{Lu22}. Similarly, gaze-based selection performance for targets floating in front of the user while walking varies depending on how the UI is stabilized. Performance may differ when only vertical movement is affixed, as in many such UI placements~\cite{Microsoft21-BodyLocking, Klose19, Lages_Bowman19}, compared to when lateral movement relative to the user’s head is also dampened~\cite{Manakhov24-GazeOnTheGo}. Raycasting with the hands is likewise influenced by UI placement. Performance differs depending on whether targets are affixed relative to the user’s movement direction or to their head, as coordinating head and hand movements during locomotion is inherently challenging~\cite{Li24-TargetSelectionOnTheMove}. Comparable effects have been observed across a variety of UI placements and interaction techniques used during locomotion (for a detailed overview, see~\cite[Section 2.1.3]{Manakhov25-PhDThesis}).

These findings demonstrate that results obtained for one UI placement rarely generalize to others. Interaction during locomotion cannot be considered in isolation from the spatial relationship between the user and the AR interface. Thus, the \textbf{research challenge} lies in our limited understanding of how the performance of interaction techniques on the go is mediated by UI placement. The relative movement between the UI and the user must be placed at the center of analysis. Interaction techniques intended for on-the-go use should be designed with their target UI placements in mind, and experimental evaluations would benefit from treating UI placement as an independent variable.

\paragraph{Conceptualisation and Design Space for UI Placement in AR}

We are yet to discover many UI placements suitable for interaction on the go. This exploration is shaped in part by how we conceptualize UI placement. For example, understanding placement in terms of the objects to which interfaces are anchored — a perspective established by early seminal works~\cite{Feiner93, Billinghurst98-SpatialConferencingSpace, Bowman04} — implicitly constrains the design space to placements where UIs are positioned at fixed offsets relative to such objects. This perspective overlooks more complex designs that may be particularly suitable for locomotion. Examples include tag-along UIs that remain stationary relative to the world within a defined threshold around the user but are pulled along once that threshold is exceeded~\cite{Lages_Bowman19-WalkingWithARWorkspaces} (\autoref{fig:nonTraditionalUIplacements}a), or UIs that travel with the user while dynamically adjusting their position to avoid overlapping with world geometry~\cite{Belo22}. Alternatively, UI placement can be conceptualized as a function determining how UIs are positioned based on contextual inputs. This view reveals nuances obscured by generic labels such as “world-referenced”. For instance, one can imagine a range of mobile world-referenced placements: from designs that periodically “push” the UI several meters ahead of the user as they walk, to systems that detect suitable surfaces (e.g., billboards) in the environment and align UIs with them (\autoref{fig:nonTraditionalUIplacements}b).

The \textbf{research opportunity}, therefore, lies in reconceptualizing UI placement in a way that broadens the design space and allows to identify specific differences in UI placements that meaningfully affect interaction on the go. Developing such a conceptualisation would not only structure the space of mobile UI placements but also support its systematic extension.

\vspace{0.2cm}
Deepening our understanding of the notion of UI placement itself and its effects on input on the go will ensure seamless interaction with future generations of wearable AR during locomotion.

\bibliographystyle{ACM-Reference-Format}
\bibliography{references}

\end{document}